\documentclass[useAMS,usenatbib]{svjour3}
\usepackage{epsfig}
\usepackage{amsmath}

\newcommand{\ups}{\rule{0pt}{15pt}}

\begin{document}

\title{The revision of the turbulence profiles restoration from MASS scintillation indices}

\author{Victor G. Kornilov \and Matwey V. Kornilov}

\institute{ V. Kornilov 
           \at Sternberg Astronomical Institute, Universitetsky prosp. 13, 119992 Moscow, Russia\\
           \email{victor@sai.msu.ru}
           \and M. Kornilov
           \at Sternberg Astronomical Institute, Universitetsky prosp. 13, 119992 Moscow, Russia \\
           \email{matwey@sai.msu.ru}
}

\date{Received: date / Accepted: date}

\maketitle
\begin{abstract}
The altitude distribution of optical turbulence is derived from the MASS instrument data by solving an inverse problem. In this paper, some modifications of the profile restoration are described. The principal change is the introduction of the Non Negative Least Squares algorithm which has good regularizing properties. An averaging of scintillation indices was replaced with averaging of obtained solutions what leads to clearer physical results. It is shown that restoration with a number of turbulent layers as large as 14--15  can be successfully performed.
\end{abstract}

\section{Introduction}
It is well known that the efficiency of astronomical observations in optical and near IR range greatly depends on turbulence in the earth's atmosphere. Modern techniques for telescope efficiency gain require both statistically-valid long-term and near real-time information about properties of the optical turbulence (OT) above an observatory. In general, the turbulence intensity is described with help of the refraction index structure constant $C_n^2$. One of the instruments designed to measure this parameter is MASS (Multi Aperture Scintillation Sensor) \cite{Marr} developed a decade ago. More than 30 such devices are used in different astronomical observatories and site testing campaigns  \cite{Kor07}.

The method used by the MASS \cite{Marr,MASS} to determine the OT vertical profile is based on the simultaneous measurements of stellar scintillations in four entrance apertures ($A$, $B$, $C$, $D$) of different size and produces four normal and six differential \cite{T98} scintillation indices $s^2$ --- a variance of the relative fluctuations of light fluxes. The measurement is performed with an integration time $T_e$ of 1~ms and an estimation of $s^2$ is calculated from the measurements at time interval $T_b$ ({\it basetime}) of about 1~s \cite{MASS,mnras2003}.

Scintillation indices $s^2$ are affected by turbulence strength in the whole atmosphere traveled through by the light:
\begin{equation}
s^2 = \int_0^\infty C_n^2(h) Q(h)\,dh,
\label{eq:sc}
\end{equation}
where $C_n^2(h)$ is index constant at altitude $h$, $Q(h)$ is certain weighting function that depends on geometry of light propagation: size of receiving aperture and angular size of emitting object. The function $Q(h)$ can be calculated for a given geometry and stated spatial spectrum of refraction index perturbation \cite{MASS,mnras2003}.

The calculation of the vertical distribution of the OT from measured $s^2$ indices is only possible by solving the inverse problem which is ill-conditioned for any practical set of MASS apertures (therefore, set of $Q(h)$ functions). For this reason, the restoration algorithm of the turbulence profiles is a key part of the MASS method. The description of the algorithm and its verification are given in paper \cite{mnras2003}. Instrumental and other factors affecting accuracy of MASS results were considered in \cite{Kor07,mnras2007}.

During development of the algorithm used in the {\it Turbina} software (hereafter --- algorithm P), some simplifications were adopted with intuitive guesses  due to lack of real data for careful analysis of the algorithm work.

In recent years, intensive measurements with the MASS instrument have been performed by different groups at different sites. The large amount of data encouraged us to revise the profiles restoration technique. The revision focuses on the implementation of a more valid mathematical standpoint and on the verification (and correction if it is necessary) of the model assumptions.

A particularly significant factor for us is the completion of a three year campaign at mount Maidanak \cite{maid2005} and a successful two year campaign at Shatdzhatmaz summit \cite{mnras2010} where 2.5~m telescope of Sternberg institute should be installed.

\section{Algorithm P of Turbina software}
\label{sec:p}
Formula (\ref{eq:sc}), which comes from the theory of weak perturbation, is linear in $C_n^2(h)$ and can be used to describe the effects of turbulence in a typical astronomical night. For the case of Kolmogorov's model, optical turbulence in the whole atmosphere or a certain altitudinal range is defined by one parameter only. Usually, Fried radius  $r_0$ (atmospheric coherence radius) is applied, but for description of OT distribution along the line of sight, the turbulence intensity $J$ is preferable
\begin{equation}
J = \int C_n^2(z)\, dz, = 0.06\cdot \lambda^2 r_0^{-5/3} \quad\mbox{or}\quad J = 1.5\cdot10^{-14}\cdot r_0^{-5/3} \quad\mbox{if } \lambda = 500\mbox{ nm}
\label{eq:J}
\end{equation}

Together with $J$, the seeing $\beta$ (in the conventional sense) is useful. Scaled to arcsec it is 
$\beta = 2\cdot10^7\cdot J^{3/5}$ for propagating light of wavelength $\lambda = 500$~nm.

To solve the OT restoration problem, the integral equations (\ref{eq:sc}) are written in discrete form for some fixed altitude grid:
${\bf h} = (h_0, h_1 \dots, h_{n-1})$:
\begin{equation}
{\rm A}{\bf x} = {\bf b},
\label{eq:1}
\end{equation}
where the vector ${\bf x} = (J_0, J_1 \dots, J_{n-1})$ is the solution of turbulence intensities, the vector ${\bf b} = (\langle s^2_0\rangle, \dots,\langle s^2_{k-1}\rangle)$ is a set of measured scintillation indices and $\rm A$ is a matrix $k\times n$ of values of atmospheric weighting functions $Q_j(h_i)$, calculated for given $j$-th apertures. 

Turbulence intensity $J_i$ is represented by the integral:
\begin{equation}
J = \int_{h}^{h+\Delta h}C_n^2(z) dz = \Delta h \langle C_n^2\rangle,
\label{eq:intens}
\end{equation}
where $\langle C_n^2\rangle$ is the mean value of the structure coefficient within the $i$-th layer.
This is not the only way to solve the problem, alternative methods will be discussed in Section~\ref{sec:models}.

The standard input values that remain unchanged throughout the process are: the number of used indices $k=10$, the number of layers $n = 6$, the altitude grid  0.5, 1, 2, 4, 8, 16~km. Input data $\langle s^2\rangle$ are mean values over accumulation time $T_a$ ({\it accumtime}) that is about 1 minute.

The similar characteristics of the weighting functions $Q_j(h)$ (e.g. Fig.~1 in \cite{Kor07})  lead to a large condition number of the matrix ${\rm A}$ (certainly $> 1000$), and a poorly bound equation system. Without an a priori information, a proper solution of Least Squares Problem (LSP) can only be obtained if the input data have small relative errors which is encountered in the case of strong turbulence.

The physical nature of the problem lets us apply additional restrictions: the solution ${\bf x}$ of the system (\ref{eq:1}) must be non-negative. To solve (\ref{eq:1}) with  the additional restriction ${\bf x} \ge {\bf 0}$, the direct minimization of weighted sum of residual squares with the help of the Powell method \cite{powel} was implemented in the procedure {\it Atmos}\footnote{see  http://curl.sai.msu.ru/mass/download/doc/dataproc.pdf}. To provide the restrictions, the original variables are replaced by  $v_i = J_i^{1/2}$, i.e. the problem becomes nonlinear and the final solution is obtained by $v_i$ squaring. 

The non-linearity of the problem (the function to be minimized becomes polynomial of 4 degree in $v_i$) leads to the appearance of a number of local minimums, where minimization process may be completed with some probability. The non-uniqueness of solutions, topical computational cost, and difficulties in estimating the solutions errors force us to find more mathematically reasonable algorithms for the OT profiles restoration.

\section{Non-negative least squares (NNLS)}
\label{sec:nnls}
A solving a system of linear equations in terms of least squares with non-negative solutions is not new. The technique was put forward 40 years ago and named NNLS (Non-Negative Least Squares) \cite{nnlse}. The algorithm is based on the fact that the minimum of the sum of the squares $R^2 = ||{\rm A}{\bf x}-{\bf b}||^2$ lies either into restricted domain or on the its border. To find the $R^2$ minimum one cannot set a negative components of the non-restricted solution to zero because the main axes of the residual paraboloid  do not coincide with the coordinates axes. It was proved that NNLS finishes in a finite number of iterations and usually requires only about n/2 iterations \cite{nnlse}. 

The NNLS technique is used in astronomical data processing, although not really frequently \cite{a2,a4,a5,a3,a6,a1}. Goodwin \cite{slod1} successfully applies NNLS for restoration of altitude turbulence profile from SLODAR measurements.

Following the book \cite{nnlse}, we implemented NNLS and other algorithms needed to solve a system of linear equations related to the LSP using pure C++ and boost (http://boost.org) libraries\footnote{LSP library is available for download http://curl.sai.msu.ru/$\sim$matwey/lsp/}. The LSP is solved by singular value decomposition (SVD) using Householder transforms and the modified QR-decomposition.

To obtain the best unbiased solution, the source set of equations (\ref{eq:1}) must be weighted through a left multiplication by matrix {\rm W}. Assuming that input errors are uncorrelated, the diagonal matrix {\rm W} (where $w_{jj} = (1/\sigma^2_j)^{1/2}$, and $\sigma^2_j$ is an estimator of variance of $j-$index error) is used like in the case of algorithm P.

The replacement of the solving algorithm was the first step of the revision. This version is referred below as N1.

\section{Comparison of the turbulence profile restoration algorithms}

In order to compare algorithms N1 and P, we used data taken with the original MASS device \cite{MASS} at mount Maidanak between 2005 and 2007 \cite{maid2005}. Both algorithms were run with identical input parameters and data set. The data consist of more than $3\cdot10^6$ seconds of measurements or 50\,000 points.

\begin{figure}[t]
\centering
\psfig{figure=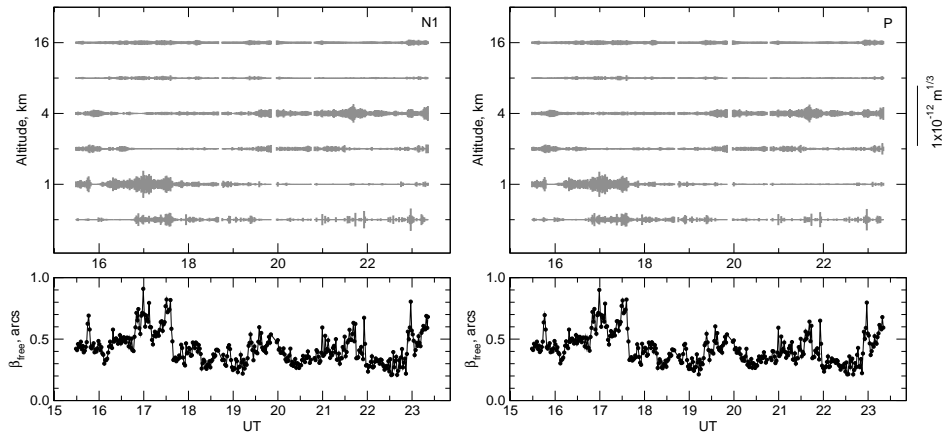,height=12.5cm,angle=-90}
\caption{ Comparison of OT profiles restored by algorithm N1 (left) and algorithm P (right) with data taken on September 6, 2005. Turbulence intensities in 0.5, 1, 2, 4, 8 and 16-km layers are shown on top, free atmosphere seeing is shown at the bottom.\label{fig:res_comp}}
\end{figure}

The principal conclusion from the comparison is that both set of solutions are in a good agreement, differing only in details. The results for September 6, 2005 are given on Fig.~\ref{fig:res_comp} as an example. One can see that all the characteristic features appear on both left and right plots. Computed with $C^2_n$ profiles  $\beta_{free}$ values are  virtually matched. The free atmosphere seeing $\beta_{free}$ \cite{mnras2003} is produced by OT above the boundary layer (conventionally above 1~km, here 0.5~km layer is included as well). The same level of agreement between the two algorithms was found on other nights regardless of turbulence intensity and distribution.

The inter-comparison of the free atmosphere seeing $\beta_{free}$ over the whole data set shows that the mean-square deviation is only $0.02''$. The systematic difference is extremely insignificant, medians $\beta_{free}$ differ less than $0.006''$ between the techniques.

\begin{figure}[t]
\centering
\psfig{figure=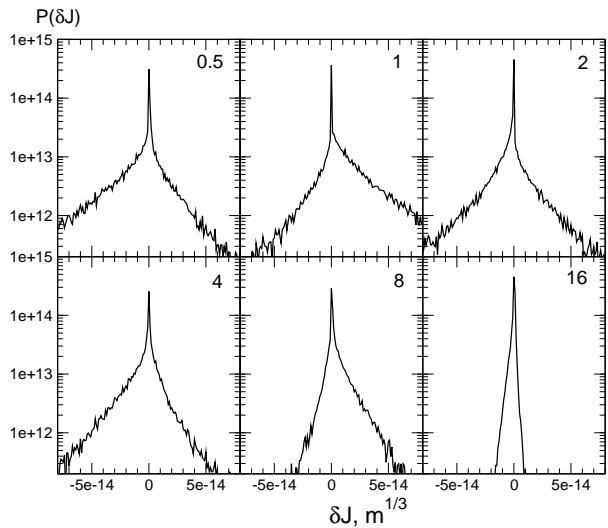,height=9cm,angle=-90}
\caption{Normalized distributions of differences  $\delta J$ for all layers. The tails of distributions look more evident due the log-scale of the vertical axis.\label{fig:diffs}}
\end{figure}

The comparison of turbulence intensities $J$ for each of the six layers occasionally  shows detectable differences. The distributions of the differences in $\delta J$ are given in Fig.~\ref{fig:diffs}. Numerical values for the distributions are listed in Table~\ref{tab:diff}. Relative fractions of $\delta J$ fallen into left and right wings are presented in the two last rows.

\begin{table}[!h]
\caption{Layer-by-layer comparison of the profile restoration using P and N1 techniques. Medians are given in units of $10^{-14}\mbox{m}^{1/3}$ \label{tab:diff}}
\bigskip
\centering
\begin{tabular}{l|rrrrrr}
\hline\hline
Layer altitude\ups, km             & 0.5   &  1   &  2   &  4   &  8   & 16\\[5pt]
\hline
Median for N1  \ups       & 1.64  & 1.33 & 0.00 & 2.43 & 2.37 & 2.36 \\
Median for P              & 2.52  & 0.52 & 0.20 & 2.91 & 1.89 & 2.42 \\
$\delta J < -2.0\cdot10^{-14}$, \%& 13.7 &  4.2 & 7.8  & 10.9 & 0.8  & 0  \\
$\delta J > +2.0\cdot10^{-14}$, \%&  8.2 & 19.5 & 6.0  & 4.0  & 6.6  & 0  \\[5pt]
\hline\hline
\end{tabular}
\end{table}

For low layers (0.5, 1 and 2~km) the base of their distribution is clearly asymmetrical. This leads to the redistribution of the turbulence power between the adjacent layers and changes the median of layer intensity. In the 4~km and 8~km layers the effect is fainter and insignificant for the 16~km layer. It is apparent that the redistribution of the turbulence energy is related to the termination of the minimization process into certain local (not global) minimum, likely into the nearest starting point.

\begin{figure}[t]
\centering
\psfig{figure=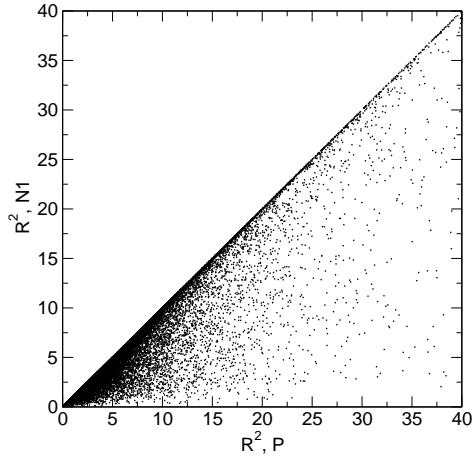,height=7cm}
\caption{Comparison of the solution residuals $R^2$ for both methods ($\approx 50\,000$ points) \label{fig:chi2}}
\end{figure}

This assumption is confirmed by Fig.~\ref{fig:chi2}, where the residuals of all the processed data are shown. Since the systems of equations are exactly the same, the cases where $R^2_{P} > R^2_{N1}$ (the points below the diagonal line) are occurrences of the algorithm P finishing in a local minimum. For instance, in 50\% of cases $R^2_{P}$ is greater than $1.06\cdot R^2_{N1}$. In a few cases, the direct minimization doesn't converge in a finite number of iterations, and the profiles of such points are therefore absent. The number of such cases is less than 0.2\%.

\section{Analysis of residuals $R^2$}

\begin{figure}[t]
\centering
\psfig{figure=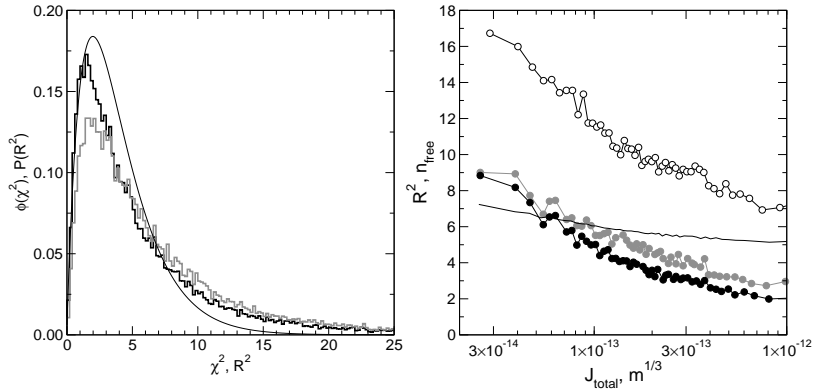,height=11cm,angle=-90}
\caption{Left: normalized distribution of $R^2$ for the N1 (black) and P algorithms (grey line). The thin line is the $\chi^2$ distribution with 4 degrees of freedom \label{fig:distr}. Right: Relationship between median $R^2$ and the total turbulence $J_{total}$ in case of algorithm P (grey points), N1 (black points) and N2 (open circles), thin line denotes the mean value of degrees of freedom $m$ \label{fig:chi2-beta}}
\end{figure}

Analysis of residuals usually helps to determine the agreement between real data and the model.  Since the $R^2$ is the weighed sum of the squares of residuals, the direct comparison with the $\chi^2$ distribution is possible. The distribution of values $R^2$ is given on the left-hand plot of Fig.~\ref{fig:distr}. The theoretical $\chi^2$ distribution with 4 degrees of freedom (10 equations with 6 variables) is also plotted. One can see that the distribution in the case of N1 is closer to the theoretical curve but a significant part of solutions do not correspond to $\chi^2$ criteria and should accordingly be dropped. In theory, $\chi^2 > 13.3$ must occur only in 1\% of events, but really it is in 10\% for N1 and 13\% for P. That is why a large threshold of order 100 was used in the output filtering of the results by values of $R^2$. The filter $R^2 > 100$ rejects 2.2\% for N1 and 3.8\% in the case of P.

Additionally, to show the adequacy of the model using $R^2$, the following conditions must be fulfilled: 1) the errors of the input data (vector ${\bf b}$) must be  estimated correctly, 2) these errors are distributed normally and almost independent, 3) the number of degrees of freedom is determined correctly.

First, the degrees of freedom are not known in advance for the linear problem with restrictions. In a sense, the restrictions are the equivalent of adding new equations to the system. The number of degrees of freedom increases when the restrictions come into effect (the solution is on the boundary of allowed area). The matrix rank on the area border is less than the rank of the source matrix. The NNLS algorithm excludes the matrix column which corresponds to a layers with zero intensity (``empty'' layer).

The dependence on the total turbulence intensity $J_{total}$ of the $R^2$ medians and the mean of degrees of freedom  $m$ (number of empty layers+4) are plotted to emphasize its importance. Medians and means were calculated using bins of 1001 points of the $J_{total}$ array. The median value of $J_{total}$ is  $1.85\cdot  10^{-13}\mbox{m}^{1/3}$, and 90 percentile is $5.41\cdot 10^{-13}\mbox{m}^{1/3}$.

These dependencies are shown on the right-hand plot of  Fig.~\ref{fig:chi2-beta}. In the case of strong OT (when in each layer the intensity is much more that error of solution) $m$ should be 4. In reality, $m$ is about 5. The number $m$ increases to 7 when the turbulence is weak, due to the appearance of ``empty'' layers.  If $R^2$ were exclusively dependent on $m$, the $R^2$ distribution would match the $m$ distribution shifted down by $\approx 2/3$.

There must therefore be additional factors that decrease $R^2$ with strong turbulence and increase it in case of weak one. The first such factor is the adequacy of the estimation of errors of input data and consequently the correctness of weighting the linear system. Obviously, that overshooting of errors leads to a decrease of $R^2$ and vice versa.

The systematic errors of the scintillation indices connected with the wrong MASS instrumental parameters or incorrect sky background subtraction or light scattering \cite{mnras2007} can be present in the input data. The influence of such errors as well as errors of weight functions $Q(h)$  (inaccuracy of matrix {\rm A}) is estimated in a different way.

\section{Properties of random errors of the input data}
\label{sect:inp-err}

The input data for the vertical profile restoration are 10 average scintillation indices $\langle s^2\rangle$ calculated  over an integration time $T_a$ from individual measurements $s^2$. The error estimates of $\langle s^2\rangle$ used for calculation of diagonal weight matrix $\rm W $ are computed in the ordinary way. The application of such weighting is valid if the input data errors are uncorrelated.

This assumption is quite valid if the input errors are originated from the photon statistics and the scintillation process itself is homogeneous on the time scale of $T_a$. To check these predicates, variances $\sigma^2_{s^2}$ and correlation coefficients $\rho_{ij}$ were calculated over sets of indices $s^2$ over $T_a$ interval. Over a small sample size ($n = T_a/T_b = 60$), the relative accuracy of the estimates is not better than 20\%. The correlation and non-normality of the scintillation indices even more degrade the accuracy.

The median errors $\sigma^2_{\langle s^2 \rangle}$ of mean indices defined as $\sigma^2_{s^2}/n$ are given in Table~\ref{tab:2}. One can see that the maximal weights will be attributed to equations for indices D, BC and CD, and  minimal weights for A and AD ones.

\begin{table}[t]
\caption{Median values of variances of errors of mean scintillation indices \label{tab:2}}
\bigskip
\centering
\begin{tabular}{c|rrrrrrrrrr}
\hline\hline
Indices\ups & $s^2_A$ & $s^2_B$ & $s^2_C$ & $s^2_D$ & $s^2_{AB}$ & $s^2_{AC}$ & $s^2_{AD}$ & $s^2_{BC}$ & $s^2_{BD}$ & $s^2_{CD}$ \\[4pt]
\hline
$\sigma^2\times10^{6}$  & 5.26 & 2.42 & 0.86 & 0.17 & 0.50 & 1.16 & 3.47 & 0.19 & 1.25 & 0.20 \ups\\[4pt]
\hline\hline
\end{tabular}
\end{table}

Examination of the correlation coefficients $\rho_{ij}$ shows that 11 of 45 median coefficients are greater than 0.8 and 21 are greater than 0.6. The correlations AB:C and AB:D are near zero which is evidence of the domination of random noise in scintillation index $s^2_{AB}$ and of weak connection via turbulent layers. On the another hand, the correlation between $s^2_A$ and $s^2_B$ is 0.88 which shows the small contribution of random noise in the indices.


It is clear that those median values correspond to situations when $J_{total}$ is close to its median value  $1.85\cdot 10^{-13}\mbox{m}^{1/3}$. Both the variances and correlation coefficients depend on the turbulence intensity. As expected, the values $\rho_{ij}$ decrease if the turbulence calms down, and hence the fraction of random fluctuation in indices is increased.

Hence, it is the turbulence non-stationarity during the time $T_a$ and not the photon noise introduces the valuable part in input data errors.  It is easy to explain because during 1 minute the atmosphere drifts by 1~km with the wind velocity is about 20 m/s what is much more than the turbulence outer scale.

It is known that, in the case of correlated input data errors, the best unbiased estimator of solution of linear system (\ref{eq:1}) is the solution of equations weighted in the following manner:
\begin{equation}
{\rm B^{-1}}{\rm A}{\bf x} = {\rm B^{-1}}{\bf b},
\label{eq:w1}
\end{equation}
where {\rm B} is the triangle matrix obtained from the decomposition of the symmetric, non-negatively defined covariance matrix ${\rm \Sigma}$ of errors of the input data ${\bf b}$. Such decomposition ${\rm \Sigma} = {\rm B B}^T$ can always be performed, for instance by the Cholesky method \cite{nnlse}.

The estimator of covariance matrix ${\rm \Sigma}$ is calculated from the same sample of indices that is used for determination of means $\langle s^2\rangle$. As before, it leads to large errors in the weight matrix ${\rm B^{-1}}$. The uncertainty of the weights affects the solution weakly, but it does give a valuable estimation of the solution errors. The Cholesky decomposition cannot be performed in some rare cases. A damping coefficient of 0.8 is used for non-diagonal matrix ${\rm \Sigma}$ elements in order to prevent such cases.

Having been weighted according (\ref{eq:w1}) the right hand side of the system  holds uncorrelated errors. 
This kind of modification (decorrelation of input errors) was implemented in the version N2 of algorithm.

\section{Estimation of random errors of the solution}
\label{sect:out-err}


An estimation of the random errors of the solution can be obtained with a general method based on Gauss-Markov theorem \cite{nnlse} by calculating the unscaled covariance matrix:
\begin{equation}
{\rm C} = ({\rm A}{\Sigma}{\rm A})^{-1}.
\end{equation}
The variance of $k$-th layer error is defined as follow
\begin{equation}
\sigma^2_i = \frac{R^2}{k-r}\cdot c_{ii},
\label{eq:vars}
\end{equation}
where $R^2$ is the residual of the solution of correctly weighted set of equations, $k$ is a number of equations (10 independent indices), $r$ is the solution rank (a number of nonzero layers). The rows and columns of the matrix {\rm C} corresponding to ``empty'' layers are identical to zero.

An estimation of correlation coefficients $\rho_{ij}$  are obtained directly from the covariance matrix
\begin{equation}
\rho_{ij} = c_{ij}/(c_{ii}c_{jj})^{1/2}.
\end{equation}

The above relationships assume the correct covariance matrix ${\Sigma}$, that is satisfied only in algorithm N2. Nevertheless, the calculation of solution errors was implemented in algorithm N1 in order to estimate an effect of applying formula (\ref{eq:w1}).

\section{Solutions and their errors for algorithms N1 and N2}
\label{sec:last}

\begin{figure}[t]
\centering
\psfig{figure=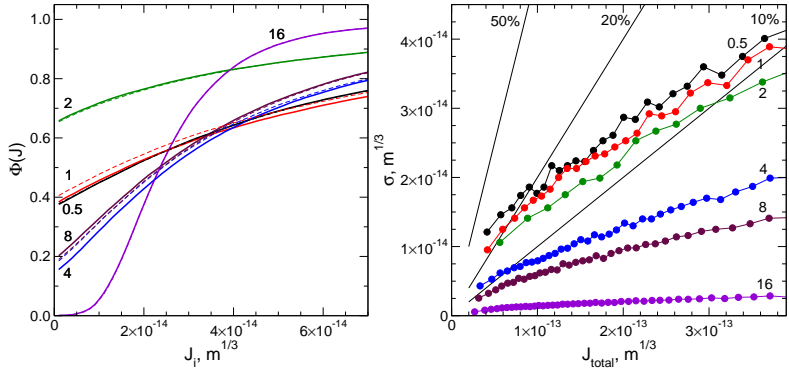,height=11cm,angle=-90}
\caption{ Left: cumulative distribution of the OT intensity in 6 layers for algorithm N2 (lines) and N1 (dashed lines). Right: dependence of medians of the solution errors on  $J_{total}$ for the N1 algorithm. The thin lines are the percentages of $J_{total}$ \label{fig:lay1}\label{fig:med-err}}
\end{figure}

The cumulative distributions of layer by layer OT obtained with algorithms N1 and N2 are given on the left-hand plot of Fig.~\ref{fig:lay1}. One can see that the curves are very close. The maximal difference is observed in 1~km layer but it doesn't exceed $3\cdot10^{-15} \mbox{m}^{1/3}$. Both curves are similar to the cumulative distribution in paper \cite{maid2005} which was obtained  with algorithm P.

The behavior of residuals $R^2$ obtained for the correctly weighted system (\ref{eq:w1}) is given on the right-hand plot of Fig.~\ref{fig:chi2-beta}. The curve is significantly higher and has a smaller inclination than one would arise from theory. The considerable increase of the residuals with decreasing $J_{total}$ could mean that the systematic discrepancy (of additive type) between the input data and the model begins to dominate. However, as it will be shown, the systematic effects are significantly lower.

For each layer, the dependence of median of the standard deviation $\sigma_i$ on  $J_{total}$ were studied in order to determine the typical errors of the restored profile. For algorithm N1 this dependence is shown on the right-hand plot of Fig.~\ref{fig:med-err}. As before, the median was calculated by the sample of 1001 points. The data related to empty layers were ignored. This method gives an estimate of the upper limit of the error.

Notice that the layers are divided into three groups: 1) the 16~km layer is determined with precision better than $5\%$ even in case of weak turbulence 2) layers 8 and 4~km are determined with precision better than 10\% when $\beta_{free} > 0.25''$ and close to 10\% in case of weaker turbulence 3) the layers 2, 1 and 0.5~km are determined with precision worse than $10\%$ when $\beta_{free} < 0.7''$ and worse than $20\%$ when $\beta_{free} < 0.3''$.

The estimations of $\sigma_i$ are slightly greater for algorithm N2 than for N1 especially in case of strong turbulence. However, they are close in case of weak turbulence. Extrapolated to zero turbulence $\sigma_i$ are respectively 1.3, 0.8, 0.7, 0.3, 0.2, 0.1 in units of $10^{-14}\mbox{m}^{1/3}$. 

For the algorithm N1, the correlations of errors $\rho_{ij}$ in the nonempty, nearest-neighbor, layers are approximately $-0.9$ to $-0.95$. For the algorithm N2, the values of $\rho$ were expected to decrease significantly but, instead, only did so by a small amount. This means that the structure of the equation set and not the characteristics of the input data is the dominant source of error correlation. The correlation of errors decreases with increasing distance between the layers. For instance, for the 2~km and 16~km layers, $\rho$ is between $+0.5$ and $-0.5$ depending on the number of empty layers between them.

The sum of the intensities of the layers is very stable thanks to the error correlation. The effect of ``dragging'' of the turbulence to the nearest-neighbor layers was marked earlier \cite{apo}. Note that such a strong correlation is the consequence of solving, and not of the physics of the phenomenon.

The question about the precision of the OT measurements in each layer was raised in \cite{mnras2003}, in the form of estimation of the noise of the restoration process. Generated by the turbulence motion over intervals less than $T_a$ errors are included in this noise and in agreement with the ones given here by order of magnitude and behavior. Our estimates coincide with the estimates of errors in \cite{apo} as well.

The problem with making valid estimates of the error in each layer is of great importance because the value of the error can be comparable with the intensity of the OT in the layer, especially for lower layers (see Fig.~\ref{fig:lay1}). It leads to significant widening of $J_i$ distribution and the underestimation of both medians and minor percentiles. The requirement of non-negativeness leads to an increase of zeros in the cumulative distribution. In some cases the median turbulence in a layer becomes zero. The underestimation of the intensity of low-altitude turbulence can significantly change the results of AO simulation. Fig.~\ref{fig:lay1} shows that only the 16~km layer has a clearly defined distribution.

In principle, knowing the estimate for $\sigma_i$, one can try to reestablish the actual distribution of turbulence using deconvolution, but the problem is not trivial because the distribution of errors depends on the turbulence in other layers.

\section{The final version: algorithm N3}
\label{sec:n3}

The significant influence of errors on the shape of the distribution rises the question: how does the distribution depend on the integration time $T_a$ and what happens during the averaging of near-neighbor data? It is clear that, in the linear problem without restrictions, the averaging of the input data over $T_a$ is equivalent to the averaging of output data over the same interval.

Detailed research of temporal behavior of the atmospheric turbulence intensity is beyond the scope of this paper. Rough estimates show that the auto-correlation function of the scintillation process can be fitted with $\sim\exp{(-t/t_0)}$, where $t_0$ is about tens of minutes that is much more than the integration time. This way, measurements of turbulence that are close in time correlate strongly (for details, see the technique of noise estimation in \cite{mnras2003}). This leads to the fact that changing the accumulation time during the profile restoration or the following averaging do not alter the results statistics, when the noise is negligible. The well-determined $J_{16}$ can be given as an example. Its distribution practically doesn't depend on $T_a$ (the difference is less than 1\%) between 5 and 240~s or by averaging 2, 4, 8 and 16 consecutive points.

For the other layers where the noise is significant, the cumulative distributions change according to integration time and averaging, owing to the non-linearity of the problem under discussion. This was the reason to change the averaging sequence in algorithm N3.

\begin{figure}[t]
\centering
\psfig{figure=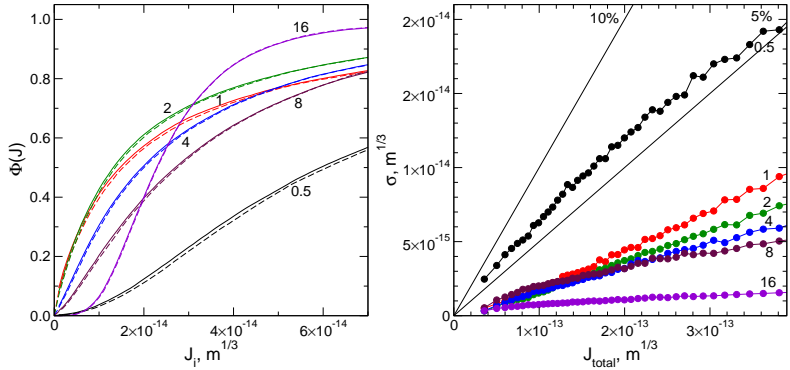,height=11cm,angle=-90}
\caption{Left: cumulative distributions of the turbulence intensity $J_{i}$ in 6 layers after restoration with 1~s intervals followed by a 1~min averaging. The distributions with a 2~min averaging are shown by the dashed lines.\label{fig:avg2} Right: dependencies of the median errors $\sigma_i$ on the total turbulence $J_{total}$ for N3 algorithm \label{fig:err3}}
\end{figure}

In the final version N3, the covariance matrix ${\rm \Sigma}$ is calculated over all the data within $T_a$ but the OT profiles are restored using individual indices $s^2$ (integration time $T_b$). The obtained profiles are averaged over an accumulation interval $T_a$ after that. Such algorithmic approach only becomes conceivable after increasing the speed of the computational processing. The cumulative distributions obtained with algorithm N3 are given on the left-hand plot of Fig.~\ref{fig:avg2}.

It can be seen that the distributions of all of layers (with the exception of the 16~km layer) have undergone significant changes relative to the curves in Fig.~\ref{fig:lay1}. A jump in zero for 0.5~km layers disappeared almost entirely. The medians increase for all of the layers, nonzero 25\% percentile appears for other layers.

The median of the turbulence in layer 0.5~km is $5.95\cdot 10^{-14}\mbox{m}^{1/3}$, that is more that 3 times the former value. The overall turbulence $J_{total}$ has increased slightly and its median value becomes $2.12\cdot 10^{-13}\mbox{m}^{1/3}$, in other words it is 13\% greater than the algorithm N1 gave, the seeing $\beta_{free}$ grew from $0.46''$ to $0.50''$.

The question about the statistical adequacy of the output set is of great importance because the distribution of the individual  solutions is extremely asymmetrical in case of weak turbulence due to the requirement of non-negativeness. Usage the median as such a characteristic leads to a turbulence distribution similar to the ones shown in Fig.~\ref{fig:lay1}. Moreover, in this case the sum of the layer intensities can differ greatly from the turbulence integrated over the whole free atmosphere.

Clearly, the sum of medians may differ from median of all the turbulence. But if this difference is large, the appropriateness of such an estimator is in question. In the case of the algorithm N1 the sum of medians is only 55\% of the total turbulence, while the algorithm N3 gives 75\%.

The usage of a mean as a central estimator leads to a physically more understandable result: the probability of zero turbulence must be zero. That estimator doesn't have to be unbiased, because the noise leads to the appearance of non-zero mean value even in the layer with zero intensity. However, such kind of bias are intrinsic for any problem with restrictions. The mean is well matched to the median in case of well-determined  turbulence and both of these estimators are unbiased in practice, for instance for high layers.

Estimates of the error of the mean over an accumulation time $T_a$ are calculated directly as root-mean-square. The behavior of the errors calculated in such a way is shown on the right-hand plot of Fig.~\ref{fig:err3}. It can be seen that the errors decrease by a factor of 3 to 5 compared to the N2 algorithm.

The accuracy of the sample variance is not high because of the amount of data in the 1~min integration, and the excess coefficient $\gamma_2$ for intensity distribution varies from 0 for the layers with fully sustained turbulence to 10 in the case where they are almost ``empty''. The relationship for the relative error of the variance $\epsilon_{\sigma ^ 2} = \sqrt{(\gamma_2+2)/N}$ shows that the accuracy of the estimator varies between 15 and 40 \%. While calculating the sample variance, the intensities are also assumed to be uncorrelated on account of their noticeable noise.

Note that in the case of N3, the criterion of the restoration quality is the average value of $R^2$ of individual solutions, which is distributed almost normally. Hence, the traditional way of detecting non-valid results is possible.

\section{Possible methods for turbulence profile discretization}
\label{sec:models}

Let ${h_i}$ be the altitude grid and $J_i$ the turbulence intensity collected (\ref{eq:intens}) between layers $h_i$ and $h_{i+1}$. Strictly speaking, we obtain the following equation from the mean-value theorem:
\begin{equation}
\Delta s^2 = \int_{h_i}^{h_{i+1}}C_n^2(h)Q(h) dh = J_i Q(\tilde h_i),
\label{eq:point}
\end{equation}
where $\tilde h_i$ is some altitude between  $h_i$ and $h_{i+1}$. Moreover, it will differ for the different indices $s_j$. The equation set (\ref{eq:1}) assumes $\tilde h_{i,j}$ are the same for all indices and matches some effective altitude $h_i^*$ to which $J_i$ is bound; for instance, a mean or geometrical mean. For this model the real layer thickness is not known exactly which doesn't lead to a problem when operating in $J_i$ space, but estimation of the corresponding $C_n^2(h)$ becomes inaccurate.

Little error is introduced if $C_n^2(h)$ and $Q(h)$ do not change rapidly inside the layer, i.e. the layer is thin: $\Delta h/h \ll 1$. The correspondence $\tilde h_i = h_i^*$ is also valid in the case where all the turbulence is located at the effective altitude. In reality, the layer thickness is comparable with its altitude, because $h_i\approx 2 h_{i-1}$ is necessary to fill all the altitude range with 6 layers. Nevertheless both $Q(h)$ and $C_n^2(h)$ can change a number of times within the layer.

The second trivial representation is based on the assumption that the turbulence  $C_n^2(h)$ is constant within the layer and $J_i = C_n^2(h_i)(h_{i+1}-h_i)$. Then
\begin{equation}
\Delta s^2 = \int_{h_i}^{h_{i+1}}C_n^2(h)Q(h) dh = \frac{J_i}{h_{i+1}-h_i}\int_{h_i}^{h_{i+1}}Q(h) dh = J_i \langle Q(h) \rangle,
\end{equation}
and the matrix of the equation set has the mean values of the weighting function in layer $\langle Q(h)\rangle$ instead of the values $Q(h^*)$.

The third possible form is piecewise linear continuous function for the structural coefficient: $C_n^2(h)(h_{i+1}-h_i) = C_n^2(h_i)(h_{i+1}-h) + C_n^2(h_{i+1})(h-h_i)$. The turbulence $C_n^2(h_N) \equiv 0$ at the top border of atmosphere ($\approx 25\div30 \mbox{ km}$). The coefficients of equation set are also calculated from the integrals of the functions $Q(h)$ and $hQ(h)$ over the layer. Unlike previous representations, a composing the system for $C_n^2(h_i)$ in grid nodes is preferred.

\begin{figure}[t]
\centering
\psfig{figure=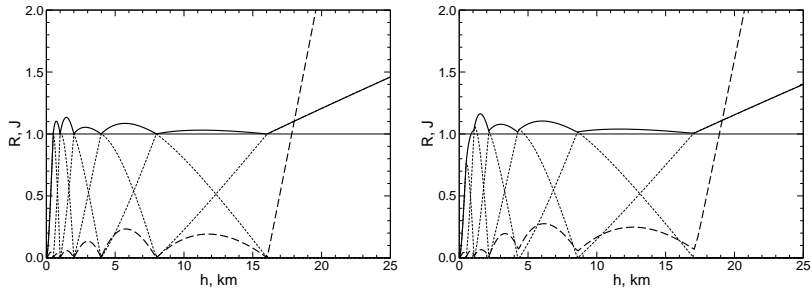,height=11cm,angle=-90}
\caption{The restoration process: contribution of the thin turbulent layers of unitary intensity at altitude $h$. The thick dashed curves are the solution residuals $\sqrt{R^2}$, the solid line is a sum over all layers. The thin dashed curves are intensities in the corresponding layer. Left: point-like representation. Right: with constant  $C_n^2$ in the layer\label{fig:res6l} }
\end{figure}

The contribution of each turbulence layer with unitary intensity located on altitude $h$ was computed for all of the described representations on the grid with nodes at 0.5, 1, 2, 4, 8 and 16~km. Similar contribution was considered in \cite{mnras2003}, but the goal of that research was to show the qualitative behavior and the possibility of the solution but not the examination from a mathematical viewpoint. The results of modeling of the first two paths are given on Fig.~\ref{fig:res6l}.

The right-hand part of the equation set corresponds to the values of the weighting functions $Q(h)$ at altitude $h$ without noise. In this case the problem is kept linear and can be solved without restriction, nevertheless, the algorithm N1 was applied (N2 and N3 were developed at a later stage).

A characteristic of the solution's residuals is evident: in the first version, the residual becomes zero when the layer altitude coincides with a node altitude. For the second and third representations, the
residual is never 0 and the altitudes of minimums do not match the nodes exactly.

The behavior of the residual changes mainly when the test layer is higher than the upper node (16~km in our case). It increases significantly and reaches 2.5 at 20~km (becomes more than 6 in sense of $R^2$). The main conclusion from this fact is that the model isn't adequate in case of existence of noticeable turbulence at altitudes above 16~km.

The restored total turbulence intensity $J$ exceeds unity by 0.05 hence the algorithm overestimate the total turbulence. All the turbulence above the upper node is assigned to it in the intensified form. For instance the turbulence at 25~km will be appended to 16~km node with a factor of 1.5. Significant turbulence at up to 25~km was noted in a number of researches. For instance, the SCIDAR measurements \cite{tiede} show the mean summer intensity above 16~km to exceed $1.5\cdot10^{-14}\mbox{m}^{1/3}$.

\begin{figure}[t]
\centering
\psfig{figure=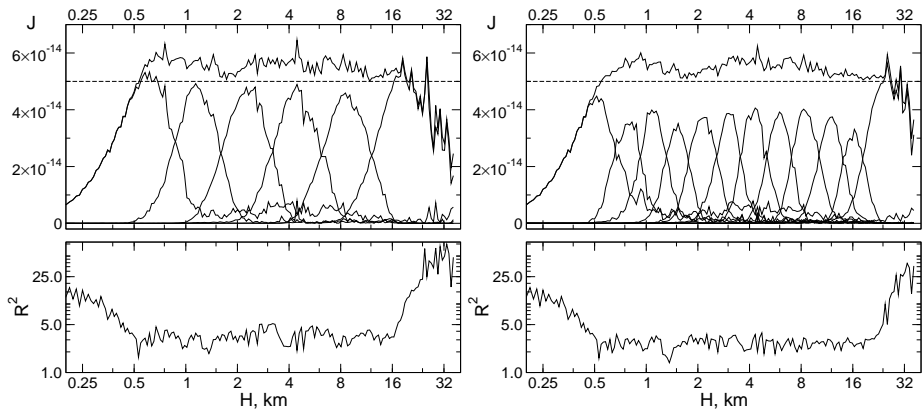,height=12.5cm,angle=-90}
\caption{Top: Reaction of the N3 algorithm to a thin turbulent layer of intensity $J = 5\cdot 10^{-14}\mbox{ m}^{1/3}$ at altitude $h$ and real errors of the scintillation indices from night of September 17, 2005. The dashed line is the expected total intensity, the thin black  is the measured total intensity. Bottom: behavior of the residual $R^2$. Left:  6-layer restoration model. Right: 12-layer restoration model \label{fig:simul12} \label{fig:simul6}}
\end{figure}

It is possible to combine the requirements of increasing the upper boundary and decreasing the layer thickness by transitioning from a 6-layer model to models with a larger number of layers. For example, by adding a lower node at an altitude close to 0.35~km and an upper node at an altitude of about 25~km in order to guarantee the absence of turbulence above the last node.

The differences between the three representations of the problem are not important and there is no need to reject the linear system of the form (\ref{eq:point}).

In the case where noise is present, the reaction of the model to thin turbulent layers is different. In order to simulate a real situation, the indices $s^2$ were normalized in such a way that their average over $T_a$ corresponds to a turbulence layer with fixed intensity at a needed altitude. The relative indices fluctuations are kept in this case. If the photon noise does not dominate then the covariance matrix and the weighting matrix are unchanged.

Simulation results are given on the left-hand plot of Fig.~\ref{fig:simul6}. The measurements of September 17, 2005 were used as noise template. One can see two significant differences from the noiseless model: 1) The residual is much closer to its average value if the layer is located below the 0.5~km node (the first node of the altitude grid) 2) The residual $R^2$ increases rapidly with increasing altitude of the layer above the upper node, but the solution continues to fall. Similar behaviors were observed for the entire night of data.

One can see that the mean value of $J_{total}$ is about 10\% over the input intensity partially due to the noise bias. Also, the intensity in the first and the last layers is slightly overestimated. The ``tail'' of the 0.5~km layer originated by upper layers noise is a characteristic detail. A dim star ($\approx 5.5\mbox{ pulse/ms}$) was acquired in the second half of the night and the intensity of the noise ``tail'' increased to about 30\% of $J_{total}$. However, if a layer is added at 0.35~km , all the noise migrates to it and the 0.5~km curve becomes normal.

\section{Study of the profile restoration with larger number of layers}

Efforts to obtain OT profiles with more than 6 layers were previously attempted with the algorithm P, but even the addition of two layers resulted in a significant growth of the solution's instability. Clearly, in the case of an unrestricted linear problem, the maximal size of the solution cannot be larger than the number of input data (10 scintillation indices in our case). The restrictions give the possibility to increase the number of grid nodes above 10. The new algorithm N3 which uses the great regularizing properties of NNLS makes it possible to use twice the number of layers without loss of solution accuracy.

\begin{figure}[t]
\centering
\psfig{figure=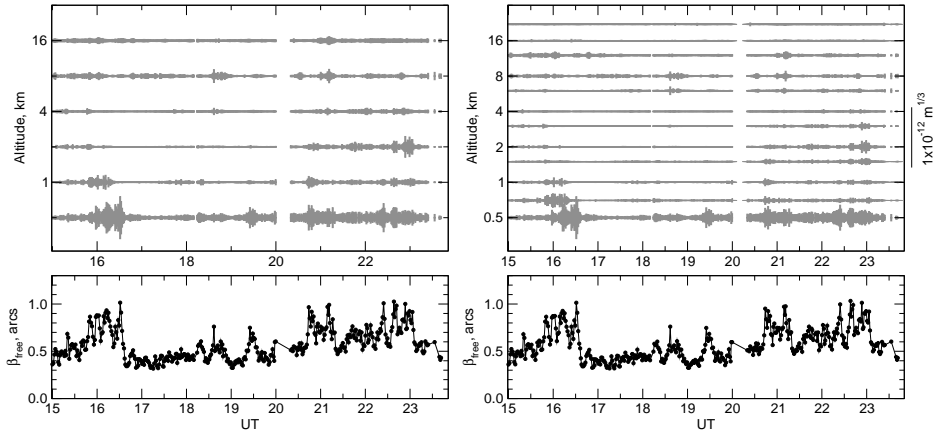,height=12.5cm,angle=-90}  \\
\caption{ Processing results of the measurements made at Maidanak on September 17 2005. Left:  6-layer restoration model. Right: 12-layer model. Bottom panels: free seeing $\beta_{free}$ \label{fig:comp6-12}}
\end{figure}

Note that the denser grid does not necessarily increase of resolution. For a real resolution increase, a low level of input ``atmospheric'' noise (stationarity of OT) is needed too. In any case, additional nodes let to localize  features of the turbulence in altitude more precisely. It is very important for understanding a turbulence generation process above a site. The study of a 12-layer restoration is given below although the experiment shows that solving with 14--15 layers is possible too.

For simplicity the 12-layer grid was obtained from the 6-layer grid by adding intermediate layers with 1.5 times the altitude. In Fig.~\ref{fig:comp6-12} the results of 6 and 12-layers OT restoration are shown. The night of September 17 2005 is calm enough, although a number of bursts are registered during the night practically in each layer. The comparison shows that high turbulence is located not at 16~km but between 8 and 12~km where the tropopause must be.

The total integrated turbulence is the same as one can see from comparison of seeings $\beta_{free}$ computed by profiles. In the 22~km layer the weak turbulence is collected, its reality may be questioned but this layer is needed to accumulate the highest turbulence. Low altitude turbulence is attributed to the 0.5~km layer. When 13-th layer at altitude 0.35~km was added, majority of 0.5~km turbulence migrates to this layer. Most likely, the turbulence observed at 0.5~km comes mainly from lower altitudes.

\begin{figure}[t]
\centering
\psfig{figure=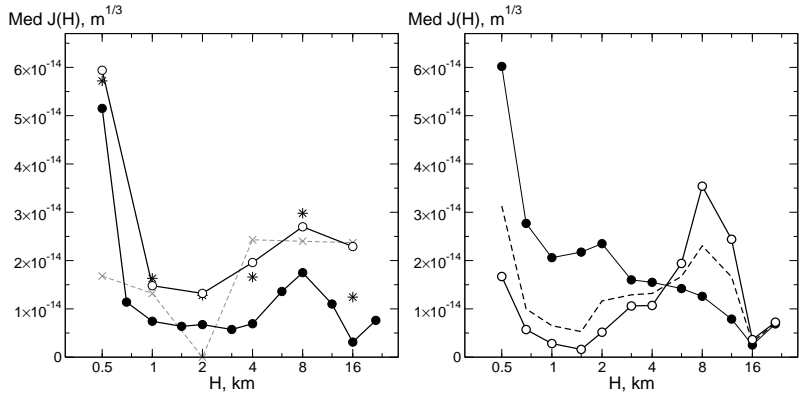,height=11cm,angle=-90}
\caption{Left: Median turbulence intensity profiles at Maidanak for both 6-layer (open circle) and 12-layer  (black circles) models. Asterisk ---  12-layer data recalculated on 6 layers for comparison, grey crosses --- data from \cite{maid2005}.\label{fig:prof2} Right: Median turbulence intensity profiles at Tolar in January 2004  (black points) and in October 2005 (open circles) obtained with 12-layers model. Dashed line --- the mean profile\label{fig:prof-tol}}
\end{figure}

Turbulence intensity profiles for 6 and 12-layers models are shown in Fig.~\ref{fig:prof2}. It can be seen that they have a physically correct behavior: strong near the ground and boundary turbulence become apparent in 0.5~km layer, next increase at 6 -- 10~km in tropopause zone and decrease at highest altitude after.

To estimate the real altitudinal resolution, the thin layer generation described above was used. As for the 6-layers model the night September 17 2005 was selected as a noise template. The simulation results are presented on the right-hand plot of Fig.~\ref{fig:simul12}. In general, the response looks similarly to the 6-layer model, however the lowest and the highest layers differ more from the others. The maximal intensity in intermediate layers is $\approx 75\%$ of the theoretical value. This indicates that the 12-layer grid is denser than the resolution for the current ``atmospheric'' noise produced by non-stationarity.

The picture is more complex for real turbulence distribution where a strong layer may be near weak one. Depending on their disposition, the weak layer turbulence flows partially either to or from the strong layer. Of course, the 12-layer model is more sensitive to this effect and to the noise than the 6-layer model.

The problem of the top layer  is solved by adding an extra node at an altitude of 23 -- 25~km, where the OT is very weak. Dealing with the bottom layer is a more complex problem. In this layer the turbulence is very strong and it therefore accumulates a lot of noise. An additional node makes it possible to specify the lower boundary of the OT. For example, in Fig.~\ref{fig:tololo14} the layer at 0.35~km  drives the turbulence higher than 0.4~km to the 0.5~km layer which is free from most fraction of the noise power.

\section{Validation of the profile restoration for MASS/DIMM device measurements}

Previous studies were performed using the original MASS instrument. However, most of the data are delivered nowadays by MASS/DIMM devices \cite{Kor07}, which have different aperture set and spectral response --- i. e. different set of atmospheric weighting functions $Q_j(h)$.

\begin{figure}[t]
\centering
\psfig{figure=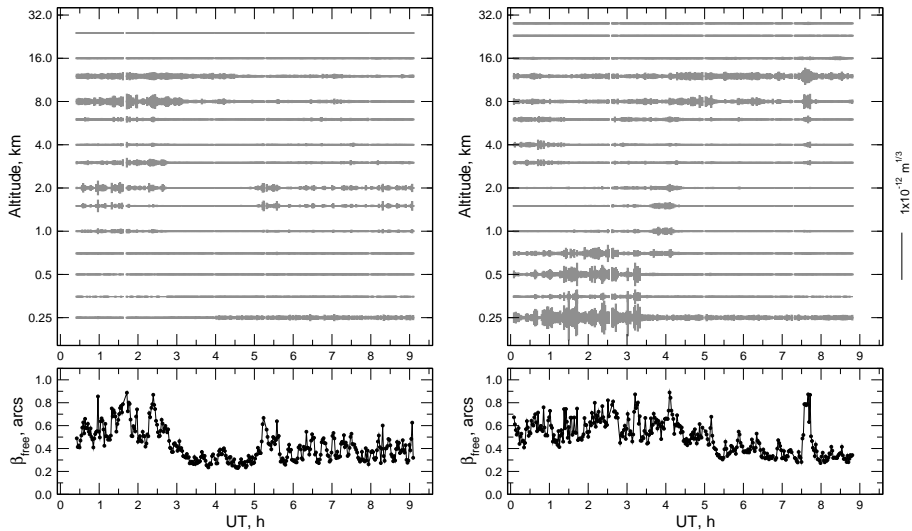,height=12.5cm,angle=-90}
\caption{Results of the OT profiles restoration for data of measurements on Cerro Tolar on October 1 (left) and October 19 (right) 2005 \label{fig:tololo14}}
\end{figure}

Although these differences are not fundamental, it was necessary to verify the final algorithm N3 for the MASS/DIMM data. The measurements at mount Cerro Tolar (Chile) carried out by the TMT group \cite{tmt1,tmt2} in January 2004 (11 nights) and October 2005 (17 nights) were used as input data. In January, the MASS/DIMM segmentator mask  produced a ghost image in the B channel direction, which resulted in large residuals in the restoration. The problem was fully fixed by accounting for scattered light as large as 0.12.

Profiles were restored using 12-layers grid. The character of the OT in these two seasons was found essentially different (see right-hand plot on Fig.~\ref{fig:prof-tol}). In Fig.~\ref{fig:tololo14} the restored OT in October 1 and October 19  nights are shown.

To detect the accumulation of ``undue'' turbulence in the lowest and highest layers, two very low layers (0.25 and 0.35~km) were added  to the 12-layers ($0.5, 0.7, 1 \dots 16, 22$~km) model. Additionally, in the processing of October 19 a very high layer 32~km was added.

One can see that the N3 restoration algorithm works perfectly for both the 14 and 15 layer grid. On October 1, residual $R^2$ is about 1 before  $4^h$ UT and $\approx2$ after.  The reason is that a bright star ($750\mbox{ pulses/ms}$ in D aperture) was changed by a fainter one ($\approx200\mbox{ pulses/ms}$) in that moment. This event clearly affects the 0.25~km layer intensities. On October 19 a bright star was used until $2^h45^m$ UT, therefore the strong 0.25~km turbulence before $3^h15^m$ is certainly real.

Measurements made with the SLODAR \cite{slod2} show that stratification is often observed in the boundary layer. So, weaker turbulence in the 0.35~km than in its neighboring layers may be real. Also, on can see that uppermost OT is not located at 16~km but in the 8 -- 12~km altitude range. Such correction increases the isoplanatic angle when calculated from the profiles.

\section{Conclusion}

The described modifications of the OT profiles restoration process do not, in principle, change the physical results obtained with MASS instrument. However, the revised restoration algorithm makes the results more reliable and clearer for interpretation. Studies of the profile restoration were carried out before in connection with MASS instrument verification \cite{mnras2003,mnras2007,apo}. Many discovered effects stimulated the work presented in this paper.

We formulate a summary of the modifications made to the algorithm and their consequences on the results obtained with the MASS technique.

\begin{enumerate}

\item Algorithm based on direct minimization is replaced by a more mathematically correct method of Non-Negative Least Squares (NNLS).

\item Comparative analysis of the restoration of altitude profiles of the optical turbulence on measurements at mount Maidanak in 2005 -- 2007 shows that the NNLS algorithm works always better.

\item Calculation of solution errors has shown that in the case of weak turbulence errors are comparable with the turbulence intensities.

\item The consequence of that is the distortion of the statistical distribution of layer intensities, leading to a significant underestimation of the median and quartile values.

\item The main source of uncertainty is the non-stationarity of OT at the scale of the integration time (usually 60~s), noise of the measurements is almost always much lower.

\item De-correlation of the input data errors slightly changes the estimates of output errors and does not change the solutions themselves.

\item The change of the strategy of the restoration algorithm from ``averaging indices -- restoration''  to ``restoration -- averaging profiles'' has led to a significant decrease of the errors of the OT profiles.

\item As a result, vertical profiles have more statistical significance and physical clarity. The integrated turbulence intensity in free atmosphere has increased by no more than 15 \%.

\item In the case of turbulence, corresponding to good seeing, it is possible to increase the vertical resolution and accuracy of turbulent layer localization (from $ h / 2 $ to $ h / 4 $) using the  12 -- 14 layers model.

\item The new algorithm is also effective for data obtained with the MASS/DIMM instrument, which has  different set of the entrance apertures.

\end{enumerate}

An additional consequence from the revision of restoration process is the improvement of the C++ code that is clearer and more suitable for further modifications and a substantial (order of magnitude) increase in processing speed.

\begin{acknowledgements}
We are grateful to the MASS team in Sternberg astronomical institute, the team that performed measurements on the mount Maidanak, the TMT group for the data from Tolar in Chile, A.\,Tokovinin (CTIO NOAO) and T.\,Travouillon (TMT) for helpful discussions and M.Sarazin (ESO) for his interest to this work.
\end{acknowledgements}

\end{document}